 \documentstyle[12pt]{article}
\def\ll{\label}
\def\re{\ref}
\def\c{\cite}

\def\r1{(\ref{$1})}
\def\ot{\otimes}

\def\ba{\begin{array}{c}}

\def\ea{\end{array}}

\def\ni{\noindent}
\def\si{\sigma}
\def\da{\dagger}
\def\De{\Delta}

\def\ov{\over}
\def\ha{{1\over 2}}

\def\l{\left}
\def\l({\left(}
\def\r){\right)}
\def\r{\right}
\def\rw{\rightarrow}

\def\la{\lambda}
\def\al{\alpha}

\def\be{\begin{equation}}
\def\bc{\begin{center}}
\def\ec{\end{center}}
\def\bit{\begin{itemize}}
\def\eit{\end{itemize}}
\def\ee{\end{equation}}
\def\ed{\end{document}}
\def\bea{\begin{eqnarray}}
\def\eea{\end{eqnarray}}
\def\efr{\end{flushright}}

\begin{document}
\title{Algebraic construction of quantum integrable models
including inhomogeneous models
}

\author{
Anjan Kundu \footnote {email: anjan@tnp.saha.ernet.in} \\  
  Saha Institute of Nuclear Physics,  
 Theory Group \\
 1/AF Bidhan Nagar, Calcutta 700 064, India.
 }
\maketitle
\vskip 1 cm

\begin{abstract} 
Exploiting the quantum integrability condition we construct 
an ancestor model associated with a new underlying quadratic
algebra. 
This ancestor model represents an exactly integrable quantum
 lattice inhomogeneous  anisotropic
 model and  at its various realizations and limits can 
generate  a wide range of  
  integrable models. They   cover quantum lattice as well as field models   
associated with the quantum  
  $R$-matrix of trigonometric type or  at the undeformed  $q \rightarrow 1$
limit   similar models belonging to  the  rational class. The classical
limit likewise yields the corresponding classical discrete and field models.
Thus 
 along with the generation of known integrable  models 	
in a unifying way a new class of
 inhomogeneous models including variable mass sine-Gordon model,
inhomogeneous Toda chain, impure spin chains etc. are constructed.

\medskip
\end{abstract}

\smallskip

  
\section{Introduction}
\setcounter{equation}{0}
Classical integrable systems in $1+1$ and
$0+1$ dimensions have been given an unifying picture through possible
reductions of the self-dual Yang-Mills equation \c{sdym}.  However, such
 success could not be achieved in the quantum case
  and there is therefore a genuine need for discovering some scheme, which
 would generate quantum integrable models (QIM) \c{kulskly} along with their
 Lax operators and $R$-matrices in an unifying way.

Such a scheme should  be general enough  to describe the lattice as well as the
field models and quantum as well as the corresponding classical models.
Similarly it should also have freedom  to switch  over from
relativistic to nonrelativistic and from anisotropic to isotropic models.
Therefore it is natural to demand such unifying ancestor model to be a
quantum (with quantum parameter $\hbar$), lattice (with lattice constant
$\Delta$) model containing a relativistic or anisotropic parameter $q$ along
with some possible inhomogeneity parameters $\{c\}$. From such a model therefore we
can produce different types of models at different limits of $\Delta, \hbar
$, $\{c\}$ and $q$ covering a wide range of integrable models. For example at
lattice constant $\Delta \rw 0$ one would generate relativistic quantum
field models like
sine-Gordon, Liouville model etc. and with further limit
  $\ q \rw 0$ we should get nonrelativistic 
field models like NLS model.  $\Delta \neq 0$ would yield the corresponding 
lattice versions and also discrete models like relativistic Toda chain,
anisotropic XXZ spin chain etc. or at  $q \rw 1$ limit, the nonrelativistic
Toda or the isotropic XXX spin chain. 
 $\hbar \rw 0$  at the same time should recover 
 the corresponding  classical
discrete or field models at appropriate limits.
$\{c\} \neq 0$ on the other hand yields a new class of inhomogeneous models.

To seek the unifying  structure underlying such a general integrable
 system, we note 
that quantum integrable systems  exhibit 
intimate connections with Hopf algebras like 
Lie algebras and their
 quantum deformations 
  \c{tarasov}-\c{frt}. Therefore, 
motivated by these facts and our earlier experience \c{construct},\c{qdnls},
 we   construct an ancestor model, which is quantum, discrete,
q-deformed and also inhomogeneous. This model is
 associated with a new Hopf
 algebra, which however   is not introduced
by hand but dictated by the
  integrability condition, i.e.  the Yang-Baxter equation itself.  Therefore
whenever the fields of the  constructed model satisfy
 the underlying algebra, its quantum
integrability  is guaranteed automatically.  The
ancestor model itself may be considered as a novel, inhomogeneous,
 exactly integrable generalized quantum lattice sine-Gordon model and turns
 out to be an excellent candidate for generating a wide range of 
integrable quantum models
   with $ 2 \times 2 $ Lax operators. The associated $R$-matrix is either
 the known trigonometric one or its corresponding rational form.
  
 Thus our scheme
 unifies a large class of  integrable quantum models by generating them in a
 systematic way through reductions of an { ancestor} model with explicit Lax
 operator realization. Note that the Lax operator together with the quantum
$R$-matrix define an integrable system completely, giving also all conserved
quantities including the Hamiltonian of the model.
To stress on the wide varieties of the well known 
quantum integrable models mentioned 
above,  we have provided a short list of them along with
 their representative Lax
operators  in the Appendix. This would be helpful for the 
ready reference and to follow
their  derivation from the single ancestor model in our scheme.

\section{The unifying algebra}

The   
unifying 
  algebra  proposed in our scheme 
is found to be a Hopf algebra. It is more general than the
  well known quantum Lie algebra and in contrast represents a
 quadratic algebra (QdA), so called because  the generators in
    the defining  algebraic relations appear in the quadratic form.
The algebra may be defined by the relations
\be
 [S^3,S^{\pm}] = \pm S^{\pm} , \ \ \ [ S^ {+}, S^{-} ] =
 \left ( M^+\sin (2 \al S^3) + {M^- } \cos
( 2 \al S^3  ) \right){1 \over \sin \al}, \quad  [M^\pm, \cdot]=0,
\ll{nlslq2a}\ee
where  $ M^\pm$ are  the central elements.  
 We show that (\ref{nlslq2a}) is  not merely a { modification} of the known 
$U_q(su(2))$ but is   
  a QdA underlying an integrable ancestor model and  is a
consequence of the quantum Yang Baxter equation (QYBE) 
\be
R(\la-\mu) L(\la)\otimes  
L(\mu)= (I \otimes L(\mu) ) \otimes ( L(\la)\otimes I) R(\la-\mu).
\ll{ybe}\ee
  We take  the associated
$4 \times 4$  quantum  $R(\la)$-matrix  
 as the well known  solution 
related to the sine-Gordon  model, 
  with  its nontrivial elements given by
  \c{kulskly}
\begin {equation} 
R^{11}_{11} = R^{22}_{22}= a(\la),
\  R^{12}_{12} = R^{21}_{21}= b(\la), \ R^{12}_{21} = R^{21}_{12}= c 
      ,    \ll{R-mat}\end {equation}
 and expressed through 
 trigonometric functions of spectral parameters as 
\be a(\la)=\sin(\la+\al), \ \ b(\la)= \sin \la , \ \ 
 c ={ \sin \al}. \ll{trm}\ee 
On the other hand, we  choose the    Lax operator of our ancestor model  as
\be
L_t^{(anc)}{(\xi)} = \left( \begin{array}{c}
  \xi{c_1^+} e^{i \al S^3}+ \xi^{-1}{c_1^-}  e^{-i \al S^3}\qquad \ \ 
2 \sin \al  S^-   \\
    \quad  
2 \sin \al  S^+    \qquad \ \  \xi{c_2^+}e^{-i \al S^3}+ 
\xi^{-1}{c_2^-}e^{i \al S^3}
          \end{array}   \right), \quad
          \xi=e^{i \alpha \la}. \ll{nlslq2} \ee
with $c^{\pm}_a$ being central to the algebra  (\re{nlslq2a}) 
 through the relation $ M^\pm=\pm   \sqrt {\pm 1} ( c^+_1c^-_2 \pm
c^-_1c^+_2 ). $ 
 The derivation of
algebra (\ref{nlslq2a}) follows directly 
from QYBE by inserting  in it the explicit forms of the Lax operator  
  (\ref{nlslq2}) and the $R$-matrix (\ref{R-mat})  
 and matching different powers of the
spectral parameter $\xi$.
Therefore algebra (\ref{nlslq2a}) or its various realizations
in effect becomes  equivalent to the 
QYBE and guarantees the quantum integrability of the model constructed on
it.
To establish  that
  (\ref{nlslq2a})  is   
a Hopf algebra 
we show that   
 the following characteristics  
known as 
the  coproduct  $\Delta(x): A \rw A \otimes A$, antipode $S:A \rw A  $
 and the
  counit $\epsilon: A \rw k, $  along with a 
 multiplication
 $ {\sl M}: A \otimes A \rw A $ and a unit $ \eta : k \rw
A $ hold for algebra    
(\re{nlslq2a}).
 All these  taken together 
defines it as a  
Hopf algebra.  For deriving these objects  
we exploit another inherent property of the QYBE, namely the  product of
two Lax operators  $L_{aj}L_{aj+1}$ as well as $L^{-1}_{aj}$ represent
 also   solutions of QYBE indicating its inherent Hopf algebra structure.
Using  Lax operator (\ref{nlslq2}) therefore
  we can derive 
 the coproducts   in
 the explicit form   
\bea
\Delta(  S^+)=c_1^{+ }e^{i \al S^3} \otimes S^+
                 +   S^+ \otimes c_2^{+ }e^{-i \al S^3} &
,&  \Delta(   S^-)= c_2^{- }e^{i \al S^3} \otimes  S^- +
   S^- \otimes c_1^{- }e^{-i \al S^3}\nonumber \\
\Delta( { S^3})= I \otimes { S^3}+S^3 \ot I
&,& 
\Delta(  c_i^{\pm })=c_i^{\pm } \otimes c_i^{\pm }. 
\ll{Detsa}\eea
	Inserting the coproducts  
in the algebraic relations (\re{nlslq2a}) 
one can prove after some easy steps
that the same algebra  is also  true for its tensor product expressed
through  (\re{Detsa}). The Hopf algebra property is a key
for obtaining the same YBE relations   again for the global object 
$ \ T(\la)=\prod_{i=1}^N L_i(\la) $:
$~~ R_{12}(\lambda - \mu)~ { T} (\lambda)~\otimes { T}(\mu )= (I\otimes 
{ T}(\mu )~ {
~  T\otimes I }(\lambda)~~ R_{12}(\lambda - \mu).$
 This in turn derives for  
$~t(\la)=tr T(\la)$ the quantum integrability condition $ [
t(\lambda),t(\mu)]=0 $.

Similarly one may derive the antipode or the 'inverse'  for the algebra as
  $$
  S (  S^-)= - (c_1^{+ })^{-1} e^{- i \al S^3 }S^- e^{i \al S^3 }(c_2^{+ }
)^{-1} ,  \quad 
  S ( S^+ )=
 - (c_2^{- })^{-1}e^{-i \al S^3 } S^+ e^{i \al S^3 }(c_1^{- })^{-1},
 $$ $$
S (c_i^{\pm })  = 
(c_i^{\pm })^{-1} ,  \quad  S (e^{\pm i \al S^3 })  = 
e^{\mp i\al S^3 },
$$
which satisfy also the same algebraic relations 
(\re{nlslq2a}). The counit  
   $ \epsilon  (c_i^{\pm }) = 1, \ \ \epsilon (e^{\pm i \al S^3 })=1, \ 
\epsilon (S^\pm) = 0 $ on the other
hand  maps the algebra to some  number identities.
For multiplication ${\sl M}$
 one can take the formal definition in the algebra, while
 the unit $\eta$ may be defined through the unital element $1$ as 
$\eta (\xi) \rw \xi 1.$ 

We also observe that 
 unlike Lie algebras or their deformations, due to the presence of
multiplicative operators $M^\pm$ in (\ref{nlslq2a}), it represents
quantum-deformation of a QdA. Since these operators have arbitrary
eigenvalues including zeros, they can not be removed by scaling 
and therefore generically (\ref{nlslq2a}) is different from the known
quantum algebra. Moreover different representations of $M'$s generate new
structure constants leading to a rich variety of deformed Lie algebras,
which are
related to different integrable systems. This fact becomes important for its
present application.
 The appearance of QdA in the basic integrable system should be rather  expected,
since the QYBE with $R$-matrix having $c$-number elements is itself a QdA.
 The notion of QdA  was  introduced first by Sklyanin
\c{sklyalg}.
\section {Generation  of models }
For constructing physical models  we have to find 
first  representations of (\ref{nlslq2a}) in physical variables
like  canonical variables $u,p$   with   commutation relations 
 $  [u,p]=i,$ or bosonic operators $\psi, \psi^\dag$ 
 commuting as $  [\psi,\psi^\dag]=1
$ or the spin variables $s^\pm,s^3.$  
One such representations may be given by 
\be
 S^3=u, \ \ \   S^+=  e^{-i p}g(u),\ \ \ 
 S^-=  g(u)e^{i p}.
\ll{ilsg}\ee
where
the operator function
\be g (u)= \left ( {\kappa  }+\sin \al (s-u) (M^+ \sin \al (u+s+1)
+{M^- } \cos \al (u+s+1 ) ) \right )^{\ha}  { 1 \ov \sin \al } \ll{g}\ee
containing free parameters $\kappa$ and $s$.
Inserting this realization in the ancestor Lax operator (\ref{nlslq2}) 
  one gets   a  novel  exactly integrable quantum
 model {\it generalizing lattice  SG model}. 
It is evident that only for hermitian $g(u)$ one gets $S^-=(S^+)^\dag$.
We show below the  remarkable fact 
that by fixing different eigenvalues of $M^\pm$
in the  same form  (\re{g}) of $g(u)$   and
 taking different limits of the  parameters
involved and at the same time choosing various realizations 
we can derive
a whole range of quantum integrable models including new models.
As an added advantage,  the  Lax operators of the models are obtained 
automatically from (\ref{nlslq2}), while the
$R$-matrix is  simply inherited.  The underlying algebras of the models
are also given by
the corresponding representations of   
the ancestor algebra  (\ref{nlslq2a}).
One of the reasons
 for the success   of this scheme is  the quadratic nature of the
 algebra  (\ref{nlslq2a})
 with   the explicit  
appearance of Casimir
operators $M^\pm$. Using this feature 
It is  possible  to   build  a 
  new class
of  models, that may be considered as the 
  inhomogeneous versions  of the  existing integrable models.
 The idea of  such
 construction is to take locally different representations for the central
elements, i.e. instead of taking their  eigenvalues as constants one
 should consider them to be site (and time) dependent functions.
This simply means that
 in the expressions for $g(u_j)$ in (\re{g}), $M^\pm$ should be considered as
 $ M^\pm_j$ and consequently   in  Lax operator
(\ref{nlslq2})  all $c'$s should be lattice indexed as $ c_{j}'$s.  Thus 
 the values of central elements may vary arbitrarily at
different lattice points leading to inhomogeneous lattice models. However
since the algebra remains the same they answer to the same quantum
$R$-matrices. Physically such inhomogeneities may be interpreted as 
impurities, varying  external fields,  incommensuration etc.

\subsection{Models belonging to trigonometric class}
Before constructing  inhomogeneous  models   we show first that 
 the known models, which
were  discovered earlier in isolation and mainly 
through quantization of the
existing classical models, can be reproduced  in a unified way from our
single
ancestor model.The equations corresponding to  
 such models along with the explicit forms of  their Lax
operators are listed in the Appendix for comparison.  
For the eigenvalues of the Casimir
operators fixed at
  $ M^-=0, M^+=1, $ as  easily seen,  
 (\ref{nlslq2a}) reduces 
  to the well known quantum algebra
 $U_q(su(2))$   \c{drinfeld} given by
\be
 [S^3,S^{\pm}] = \pm S^{\pm} , [ S^ {+}, S^{-} ] =  [{2} S^3]_q.
\ll{slq2}\ee
Now the simplest representation ${\vec S} = 
\ha  {\vec \si}$ derives  the integrable $XXZ$ {\it spin chain} \c{fad95},
from  (\ref{nlslq2}). On the other hand,
representation   (\ref{ilsg}) with the corresponding reduction of  (\re{g})
as
$ g (u)={1 \ov 2 \sin \al}
  \left [ 1+ \cos \al (2 u+1)
 \right ]^\ha $
recovers the quantum exact 
   {\it lattice sine-Gordon} model \c{korepinsg} with
its  Lax operator 
  obtained directly  from (\ref{nlslq2}) with all $c'$s
$=1$ (
which is compatible with $ M^-=0, M^+=1$).

 Another  unusual Lie algebra  can  be
generated from (\ref{nlslq2a}) 
  by
fixing the eigenvalues of $c'$s as 
$ \ c^+_1=c^-_2=1, \ \ \ c^-_1=c^+_2=0 
$
  which correspond to the values
  $ M^\pm=\pm \sqrt {\pm 1}.$
  This gives  an { exponentially} deformed  Lie
 algebra \be
[ S^+, S^-]= {e^{2i\al S^3} \over 2 i \sin \al }. \ll{expa} \ee
and 
reduces (\re{g}) to 
$  g(u)= {(1+e^{i \al(2 u+1)})^\ha \ov \sqrt {2} \sin \al} $. 
The Lax operator (\ref{nlslq2})
  with these values of $c'$s and the explicit form of 
$g(u)$ clearly yields the {\it exact lattice version of the 
quantum Liouville } model \c{llm}.
 Note that  the present  values of  $ M^\pm$
  may be achieved even with 
$ \ c^-_1 \not =0,$ giving  
 the same  algebra (since (\ref{nlslq2a}) depends only on $M^\pm$ )
and hence the same realization. However, the Lax operator
which depends explicitly on $c$'s gets changed reducing 
(\ref{nlslq2}) to another nontrivial structure. This is an interesting
possibility of constructing systematically 
different useful Lax operators for the same
model. For example, the present construction of
 the {\it second Liouville Lax
operator} recovers easily that of  
\c{fadliu}, invented in an involved way 
 for its Bethe ansatz solution.
On the other hand for a bit asymmetric 
choice of
eigenvalues:
$ \ \ c^+_1=c^+_2=1, \ c^-_1= -{iq }  , \ c^-_2=  {i \over  q} 
$
leading to  $ M^+=2 {\sin \al } 
, \ M^-= 2i {\cos \al }  $ we may use another
 realization 
\be S^+= -\kappa A, \ S^-= \kappa A^\dag, \ S^3= -N, \ \
 \kappa= -i (\cot \al)^\ha
,\ll{sqosl}\ee 
 which reduces  (\ref{nlslq2a}) directly  to the 
 well known 
$q$-oscillator algebra
\c{qoscl,ng}:
\be
 [A,N] = A, \ \  [A^\dag,N] =- A^ \dag,\ \ \  [ A, A^ \dag ] = {\cos (\al (2N+1))
\ov \cos \al} .
\ll{qola}\ee
Note that such direct derivation of $q$-oscillator algebra is not possible 
to obtain  from  the quantum algebra  
$U_q(su(2)).$ Another interesting point is that,  the coproduct
structure 
(\re{Detsa}) of the Casimir operators $
\Delta(  c_i^{\pm })=c_i^{\pm } \otimes c_i^{\pm } $ clearly contradicts 
the present values of  $c^-_a, a=1,2, $ which leads  to the 
$q$-oscillator algebra (\re{qola}). This  provides perhaps 
 an unexpected argument why
$q$-oscillator algebra lacks the Hopf algebra structure. Thus we have 
constructed a new quantum integrable
 {\it $q$-oscillator  model}, from which  a more physical model
can also be obtained.  Using the realization
 (\re{ilsg}) with (\re{g})   simplified as $g^2(u)=  
{[-2u]_q}$ and  the mapping 
to the  bosonic operators:
$
\psi=e^{-i p} ({
 (s-u)})^{\ha},N=s-u,
$ 
 with $ [\psi, \psi^\dagger] = 1,$ we can express
 the 
$q$-oscillator through standard bosons as 
  $ A =
     ~\psi (\frac{[2N]_q}{2N \cos \al})^{\ha}, \ N=
     \psi^\da\psi \ . $ This  converts  
the  integrable $q$-oscillator model into a bosonic model representing 
 an exact lattice version  
 of the quantum {\it derivative nonlinear Schr\"odinger equation} (QDNLS)
\c{construct} with (\ref{nlslq2}) yielding the corresponding Lax operator.
The QDNLS
 was shown to be  related to the interacting bose gas with derivative
$\delta$-function potential \c{shirman}. Fusing two such models one can
further create an integrable
 {\it massive Thirring}
 model described in \c{kulskly}.

Since our quadratic algebra allows also trivial
eigenvalues for $M^\pm$, we may  choose even
   $M^\pm=0.$ Note that this case may  be achieved with
  different sets of eigenvalues for  $c$  as
$ i) \ \ c_a^+=1 \ , a=1,2, \ \mbox{or} \ \
 ii)\ \  c_1^\mp=\pm 1  , \  \ \mbox{or} \ \  iii)\ \  c_1^+=1, \
$
with   the rest of  $c'$s being zeros.   
All these sets of values 
 lead to the same  underlying  algebra
\be
[ S^+, S^-]= 0, \ [ S^3, S^\pm]= \pm S^\pm .\ll{nula} \ee
However, they may generate   different   
  Lax operators  from (\ref{nlslq2}) due to its explicit dependence on  $c$.
  In particular, case i) leads to
  the {\it light-cone SG} model, while
 ii) and iii)   give two different  Lax operators  found in
\c{rtoda} and \c{hikami} for  the same  { relativistic 
 Toda chain}. We  have also  seen above  such
 examples
 of constructing different convenient 
Lax operators for the same  Liouville model.
In the present case  (\re{g}) gives 
simply  $g(u)=$const., therfore
 interchanging canonically  $u \rw -ip, p\rw -i u,$
  (\ref{ilsg}) yields  
\be
 S^3 =-ip, \ S^\pm=  \al e^{\mp u } \ \ 
 \ll{dtl1} \ee
 and the ancestor Lax operator  
 generates   evidently the  {\it discrete-time or relativistic
quantum Toda chain}.

\subsection{Models belonging to the rational class}
For covering a wide range of models trough reductions
 of our  ancestor model various free parameters are inbuilt in it.
 One of them is
  the deformation parameter  $q=e^\al$, which was kept generic for the above
 trigonometric class.
 Now we consider   the undeformed 
limit  $q \rw 1$ or  equivalently  $\al \rw 0$ 
 for generating the  isotropic or the nonrelativistic models
belonging to the  rational class
 and examine how the structure of the main
objects   in our scheme gets suitably modified.
 It is immediate that for  
 the existence of such  a  limit  the central elements 
 along with the generators must  be $\al $ dependent.
A consistent procedure leads to $S^\pm \rw  is^\pm, 
M^+ \rw -m^+, M^- \rw -\al
m^-, \xi \rw 1+ i \la $ reducing 
  (\ref{nlslq2a}) to 
the  algebraic relations 
\be  [ s^+ , s^- ]
=  2m^+ s^3 +m^-,\ \ \ \ 
  ~ [s^3, s^\pm]  = \pm s^\pm 
  \ll{k-alg} \ee
with $m^+=c_1^0c_2^0,\ \  m^-= c_1^1c_2^0+c_1^0c_2^1$ as
the new central elements. We see   again that it is  not a Lie but a
 QdA, since the multiplicative   operators
  $m^\pm$ can not be removed in general due to their  allowed  zero
eigenvalues. Moreover,
 algebra  (\ref{k-alg}) is also a Hopf algebra with the coproduct
 structure
\bea
\Delta(  s^+)&=&c_1^{0 }\otimes s^+
                 +   s^+ \otimes c_2^{0 } 
, \ \ \Delta(   s^-)= c_2^{0 } \otimes  s^- +
   s^- \otimes c_1^{0 }, \ \ \Delta(  s^3)=I\otimes s^3
                 +   s^3 \otimes I \\ \nonumber 
\Delta(  c_i^{0})&=&c_i^{0 } \otimes c_i^{0 } , \ \ \  
\Delta(  c_i^{1})=c_i^{0 } \otimes c_i^{1 }+c_i^{1 } \otimes c_i^{0 } 
\ll{DetsaK}\eea
giving $\Delta(  m^{+})= m^{+}\otimes  m^{+}, 
\Delta(  m^{-})= m^{+}\otimes  m^{-} +m^{-}\otimes  m^{+}.$
 An unusual feature of (\ref{k-alg})  can be observed from (\re{DetsaK}),
 that though  being 
 an undeformed
algebra it is noncocommutative in nature.

The ancestor Lax operator   (\ref{nlslq2}) at
the limit  $\al \rw 0$ is converted naturally into 
 \be
L_r{(\la)} = \left( \begin{array}{c}
 {c_1^0} (\la + {s^3})+ {c_1^1} \ \ \quad 
  s^-   \\
    \quad  
s^+    \quad \ \ 
c_2^0 (\la - {s^3})- {c_2^1}
          \end{array}   \right), \ll{LK} \ee
 with  rational dependence on spectral parameter $\la$ and 
   the quantum
 $R$-matrix (\re{R-mat}) is reduced  also to its rational form with 
\be a(\la)=
\la+\al, \ \ b(\la)= \la , \ \ \  c =\al, \ll{rrm}\ee 
 well known for the
NLS model \c{kulskly}.
Therefore the integrable systems associated with  algebra (\ref{k-alg}) and
generated by ancestor model (\ref{LK})
 would belong now to a different class, namely the rational class, all sharing
the same rational $R$-matrix (\re{rrm}). The corresponding models, a few
examples of which are listed in the appendix, are of nonrelativistic or the
isotropic type.  

It is interesting to find that at  $\al \rw 0$ the operator function
  (\re{g}) after putting $\kappa=0$ gives
$g_0(u)=i ( (s-u)(m^+ (u+s+1)+m^-))^\ha,$ which 
  using  the
 inter-bosonic map reduces the  representation
 (\ref{ilsg})  into a generalized Holstein-Primakov transformation (HPT)
 \be
 s^3=s-N, \ \ \ \    s^+= g_0(N) \psi, \ \ \  s^-= \psi^\dag g_0(N)
, \ \ \ \ g_0^2(N)=m^-+m^+ (2s -N), \ \ N=\psi^\dag \psi.
\ll{ilnls} \ee
It can be checked  to be  an exact realization of  (\ref{k-alg})
associated with the Lax operator (\ref{LK}) and
   therefore may be considered as
 an  integrable {\it generalized  lattice NLS} model. This may 
serve now
 as an  ancestor model for the rational class.
 For generating first the standard homogeneous models 
we choose 
constant eigenvalues for   the Casimir operators. One such choice
 $  m^+  =  1,m^-  =  0, $
  reduces (\ref{k-alg})   clearly to the 
   $su(2)$ algebra, which   for  spin $\ha$
representation gives the  $XXX$ {\it  spin chain} \c{dvega0} from the 
Lax operator (\re{LK}).
Bosonic realization (\ref{ilnls}) in this case  simplifies 
 to the standard  HPT
with  $ \ \ g_0^2(N)= (2s -\psi^\dag \psi) $,
  reproducing 
the known {\it lattice   NLS} model 
\c{korepinsg} from (\ref{LK}).
 
A complementary  choice
 $  m^+  =  0,m^-  =  1, $ on the other hand converts
(\re{ilnls}) simply to
$s^ {+}=\psi, s^ {-}=\psi^\dag, s^ {3} = s -N $
due to  $g_0(N)=1$ and reduces
  (\ref{k-alg}) directly to the oscillator algebra 
\be
 [\psi ,N] = \psi,\ \  [\psi^\dag,N] = -\psi^ \dag,\ \ \ \ 
 [ \psi, \psi^ \dag ] = 1.
\ll{ola}\ee
 (\ref{LK}) with this realization 
  generates  another {\it simple   
lattice NLS} model
\c{kunrag}. 
Remarkably, the trivial choice $m^\pm  = 0$  gives again 
  algebra (\ref{nula}) and   therefore the same
  realization  (\ref{dtl1}) found  for
 the relativistic case can  also be used   
    for the   {\it nonrelativistic
 Toda chain} \c{kulskly}. The related  Lax operator 
  associated  with    the 
 rational  $R$-matrix, however should   be obtained   from (\ref{LK}). 
It should be noted that a bosonic realization  of 
  general Lax operators like (\ref{nlslq2}) and
({\ref{LK}) can be found also   in  some earlier works  \c{tarasov,korl}.
\section{Field models and classical models}
As we required, our ancestor model apart from the discrete quantum 
systems obtained
above, is capable also of constructing the integrable family of quantum
field models as well as classical discrete and field models at different
limits. For obtaining the quantum field models we have to  start 
    from their respective  lattice versions constructed above and    
scale   the lattice  operators 
$ p_j, u_j , c^\pm_a, 
 \psi_j,$ 
 consistently  by the lattice spacing $\Delta$.  At the
continuum limit
$\Delta \rw 0$ one 
would obtain the field operators $p_j \rw p(x), \psi_j \rw \psi(x)$ etc.
along with the commutation relations like   $ [\psi_j,\psi_k^\dag]={\delta_{jk}
 \ov \Delta} \ \rw \
 [ \psi(x),\psi(y)]=\delta(x-y)$. 
The corresponding Lax operator ${\cal L}(x, \la)$
 for the continuum model is obtained
from its discrete counterpart as $L_j(\la) \rw I+ i\De {\cal L}(x, \la) +O(\De^2).
$ The
associated $R$-matrix however remains the same, since it does not contain
lattice constant $\De$. Thus {\it integrable field models} like sine-Gordon,
 Liouville, NLS or the derivative NLS models are obtained (see appendix) 
 from their
discrete variants constructed above.

At the classical limit $\hbar \rw 0$, all the field operators are converted
into ordinary functions with their commutators reducing to the Poisson
brackets.  Note that the $\hbar$
enters the $R$-matrix as the scaling factor $\hbar \al$ into its elements,
which in turn determine the structure constants of the 
algebraic relations.
Therefore the ancestor algebras  
(\ref{nlslq2a}) and (\ref{k-alg}) are converted into the Poisson algebras
and the $R$-matrix to its classical form: $R(\la)=I +\hbar(\la)
+O(\hbar^2).$
However our basic scheme remains almost the same.
 Remarkably,  the quantum parameter does not enter the Lax operators 
explicitly therefore the form of     
the ancestor Lax operators   
(\ref{nlslq2}) and 
(\re{LK}) remains in the same form also in the classical limit. Therefore 
we can use all the above reductions of the ancestor models for generating 
the classical model with the same success.
\section{Inhomogeneous  models}

As mentioned above the ancestor model 
(\ref{nlslq2}) and its undeformed variant
(\re{LK})
containing nontrivial Casimir operators can be used for constructing 
a new class of  integrable inhomogeneous models.  
For this the  eigenvalues of the Casimir operators 
 should  be chosen as site dependent,
 and in general time dependent functions.

   Notice that in the sine-Gordon model unlike its coupling constant the
mass parameter enters through the Casimir operator of the underlying
algebra. Therefore taking
$ M_j^+=- (\De m_j)^2,$ one can construct a variable mass (in general 
 time
dependent) discrete
SG model without spoiling its integrability.
 In the continuum  limit it would result a novel 
  {\it variable mass sine-Gordon field
 model}
  with
 the Hamiltonian 
\be{\cal H}= \int dx \left [ m(x,t) (u_t)^2 + (1/m(x,t)) (u_x)^2 +
 8(m_0-m(x,t)
\cos (2 \al u )) \right], \ll{msg1}\ee  which may arise also in physical situations
\c{msg}. 

{\it Inhomogeneous lattice NLS} model can   be obtained  by
considering site-dependent values for central elements in (\ref{LK}) and in
the
  generalized HPT (\ref{ilnls}).
 As a possible quantum field model it may correspond to
equations like {\it cylindrical NLS}
 \c{rl} with explicit coordinate dependent coefficients. In a similar way
 inhomogeneous versions of Liouville model, (non)relativistic 
Toda  etc. can be constructed. For example, taking 
$ c^a_{1} \rw c^a_{j} $ in  nonrelativistic Toda chain  we can get  a new
 integrable {\it inhomogeneous quantum  Toda chain } with the
 Hamiltonian
 \be \ H= \sum_j (p_j +{c^1_j \ov
 c^0_j})^2+{1 \ov c^0_jc^0_{j+1}} e^{u_j-u_{j+1}}. 
\ll{todah}\ee
Note that such  inhomogeneities  can not be removed through gauge
transformation or variable change,
if the inhomogeneity $c^a_j$ are time-dependent functions.

Another way of 
constructing  inhomogeneous models 
 is to  use different realizations of
 algebras (\ref{nlslq2a}) or (\ref{k-alg}) at different
lattice sites,
depending on the type of  $R$-matrix. This may lead even to different
underlying algebras and hence different Lax operators at differing sites
opening up possibilities of building various exotic inhomogeneous
 models. For example, it may be possible to construct a hybrid
models of sine-Gordon and Liouville fields, where for $x \geq 0
$ it would follow
the sine-Gordon dynamics, while for $x <0$ the Liouville dynamics. Such 
different possibilities of model constructions will be dealt elsewhere.

  Similarly   nontrivial  examples of impurity   $XXZ$ (or $XXX$) spin
chains can be obtained,
 if we replace its standard Lax operator at a single impurity 
site $m$  by  the ancestor model (\re{nlslq2}) (or (\re{LK}))
 itself or any of its representations (in spin, q-oscillator or boson).
This would give rise to  
   integrable quantum spin chains {\it with  impurity} of various nature,
which might have physical significance.   
Such a spin  Hamiltonian with impurity would look like
\[H=- ( \sum_{j \not = m,m-1}L^{xxx}(0)^{'}_{j j+1}
(L^{xxx}(0))^{-1}_{jj+1} +L^{d}(0)^{'}_{m m+1}L^{b}(d))^{-1}_{m m+1}\]\[+
L^{d}(0)_{m m-1}(L^{xxx}(0)^{'}_{m+1 m-1}L^{xxx}(0)_{m+1 m-1}^{-1})
L^{d}(0)^{-1}_{m m-1})
\]
where $L^{xxx}(\la)$ and $L^{d}(\la)$ correspond to the spin and the
impurity Lax operators, respectively.
Such a 
   Lax operator with bosonic impurity  may be derived from 
  (\re{LK}) as
 \be
L^b_m{(\la)} = \left( \begin{array}{c}
  (\la - N_m )+ { 3 \ov 2} \ \ \quad 
  a_m^\dag  \\
    \quad  
a_m  \quad \ \ 
- {1}
          \end{array}   \right). \ll{Lboson} \ee

\section {Concluding remarks}

Thus we have    prescribed  an unifying scheme for 
 constructing  
integrable systems, which covers quantum lattice as well as field models
of both relativistic or anisotropic (with $q\neq 1$) and nonrelativistic or
isotropic models (with $q=1$) 
 along with their corresponding classical counterparts. 
 Such models can be   generated systematically
 from a single ancestor model with  underlying algebra
(\ref{nlslq2a}). In general using the freedom of choosing the eigenvalues of
the Casimir operators appearing in the algebra  to be 
site as well as time dependent functions, 
we obtain inhomogeneous quantum integrable
models constituting a new class.   
 The Lax operators of the descendant  models are constructed 
from (\ref{nlslq2}) or its $q \rw  1 $ limit (\ref{LK}), while the 
variety of their concrete representations 
  are  obtained from the same  
 general form (\ref{ilsg})  at different realizations.
 The corresponding underlying algebraic structures
are the allowed reductions of (\ref{nlslq2a}). The associated quantum
$R$-matrix of the descendant models  however remains the same
   trigonometric  or the rational form  as
inherited from the ancestor model.
  This answers to the
mystery why a wide range  of integrable  models are found to share the same
$R$-matrices.
 This fact also shows
an intimate relationship among the descendant
models  inspite of their seemingly wide external differences 
and reveals
 a universal 
character for solving these models through
 algebraic Bethe ansatz (ABA) \c{kulskly,fadrev}, where the elements of the
associated $R$-matrix plays  the major role.

\section {Appendix:  Well known examples of quantum integrable models
and their Lax operators}
\setcounter{equation}{0}
\subsection { Models associated with trigonometric $R$-matrix}
\noindent
i) {\it Field models}\\ 
1. Sine-Gordon  model \begin{equation}
u_{tt}- u_{xx} = \frac {m^2}{\eta} \sin(\eta u),\qquad 
  {\cal L} = \left( \begin{array}{c} ip , \qquad
  m  \sin (\la-\eta u) \\
   m  \sin (\la+\eta u),  \qquad -ip
    \end{array} \right).
\ll{sg}\end{equation}
          \noindent
2. Liouville model  \be 
u_{tt}- u_{xx} = \frac {1}{2} e^{2\eta u}, \qquad
  {\cal L}  = i\left( \begin{array}{c} p , \quad
   \xi e^{\eta u} \\
  \frac {1}{\xi}e^{\eta u},  \quad -p
    \end{array} \right).
\ll{lm}\end{equation}
                     \ni
3. A derivative NLS (DNLS) model  
\be
i\psi_{t}- \psi_{xx} + 4i (\psi^\da\psi\psi_x=0,
, \ \   {\cal L}  = i\left( \begin{array}{c}{
  -\frac{1}{4} \xi^2+k_-\psi^\da \psi} , \
   {\xi \psi^\da} \\
  {\xi\psi},  \
  {\frac{1}{4} \xi^2-k_+\psi^\da \psi}
    \end{array} \right).
\ll{dnls}\end{equation}
\noindent
 ii) {\it Lattice Models}  \\ 
1. Anisotropic $XXZ$ spin chain  
\bea
{\cal H} = \sum_n^N\si_n^1 \si_{n+1}^1+\si_n^2 \si_{n+1}^2 +\cos \eta
\si_n^3 \si_{n+1}^3,  \ \
{L_n }(\xi) = \sin(\la + \eta \si^3 \si_{n}^3)+\sin \eta \
 ( \si^+ \si_{n}^-
+\si^- \si_{n}^+ )
\ll{XXZ}\eea
3. Lattice SG model
\begin{equation}
  L_{n}(\xi)  =
  \left( \begin{array}{c}  g(u_n)~ e^{ip_n \Delta },
 \qquad  m\Delta  \sin (\la-\eta u_n) \\
   m\Delta  \sin (\la+\eta u_n),\qquad   e^{-ip_n \Delta }~ g(u_n)
    \end{array} \right), \quad g(u_n)^2= 1 +  m^2  \Delta^2
\cos  { \eta (2u_n+{1 }) } 
\ll{L-sg}\end{equation}
\ni
4. Lattice Liouville model  
\begin{equation}
  L_{n}(\xi)  = \left( \begin{array}{c}   e^{p_n \Delta }~f(u_n)~,
 \qquad  {\Delta}{\xi}e^{\eta u_n} \\
 \frac{\Delta}{\xi}e^{\eta u_n}   ,\qquad f(u_n)~  e^{-p_n \Delta }
    \end{array} \right), \qquad f(u_n)^2=
     1 + {\Delta^2} e^{\eta(2u_n+i)}
\ll{Llm}\end{equation}
\ni
5.  Lattice  DNLS  model (A $q$-oscillator model)
\begin{equation}
  L_{n}(\xi)  =
  \left( \begin{array}{c}
  \frac{1}{\xi}q^{-N_n}- \frac{i \xi\Delta}{4}~q^{N_n+1} ,
 \qquad  {\kappa}A^\da_n \\
  {\kappa}A_n    ,\qquad
  \frac{1}{\xi}q^{N_n}+ \frac{i \xi\Delta}{4}~q^{-(N_n+1)} 
    \end{array} \right), \qquad
\ll{Ldnls}\end{equation}
\ni
7. Relativistic or discrete-time quantum  Toda chain 
\begin {equation}
H=\sum_i\left(\cosh 2\eta p_i
+\eta^2 \cosh \eta (p_i+p_{i+1})e^{(q_i-q_{i+1})}
\right), \ \ 
L_n(\xi) = \left( \begin{array}{c}
  \frac {1}{\xi}e^{\eta p_n}
-\xi  e^{-\eta p_n}, \ \ \  \eta e^{q_n}
 \\-\eta e^{-\eta q_n
}
 ,\qquad  0
          \end{array}   \right).
\ll{rtodal}\end {equation}
\subsection{ Models associated with rational $R$-matrix
 }
\noindent
i){\it Field models:} \\
 1. Nonlinear Schr\"odinger equation (NLS)
\begin{equation}
i\psi_{t}+ \psi_{xx} + \eta^2 (\psi^\da\psi)\psi=0,
      \quad
{\cal L}(\la)  = \left( \begin{array}{c} \la ,
\quad \eta \psi \\
\eta \psi^\da, \quad -\la
    \end{array} \right).
\ll{nls}\end{equation}
\noindent
 ii) {\it Lattice Models:}   \\
1. Isotropic $XXX$ spin chain 
\be
{\cal H}= \sum_n^N \si_n^1 \si_{n+1}^1+\si_n^2 \si_{n+1}^2 +
\si_n^3 \si_{n+1}^3,  \ \
L_n(\la) = i( \la {\bf I} + \eta( \si^3 \si_{n}^3+
\si^+ \si_n^-
+\si^- \si_n^+))
\ll{XXX}\ee
\ni
2. Lattice NLS  model  
\be L_n(\la) = \left( \begin{array}{c}
\la +s- \De \psi^\da \psi   \qquad    \De^\ha(2 s- \De\psi^\da
\psi)^\ha\psi^\da
 \\
\De^\ha\psi (2 s- \De\psi^\da \psi)^\ha  \qquad \quad \la -s +\De \psi^\da \psi
          \end{array}   \right).
\ll{lnls}\end {equation}
\ni
3. Toda chain    (nonrelativistic) 
\be
H=\sum_i\left(\ha p^2_i
+ e^{(q_i-q_{i+1})}
\right),\qquad
L_n(\la) = \left( \begin{array}{c}
  p_n
-\la \qquad    e^{q_n}
 \\- e^{-q_n
}
 \qquad \quad 0
          \end{array}   \right).
\ll{toda}\end {equation}

\medskip

\ni {\bf Acknowledgment}: 

I thank the
Humboldt Foundation, Germany and the organizers of ROMP99 for financial
support. 

 \end{document}